\documentclass{ws-mpla}

\usepackage[super]{cite}
\usepackage{xcolor}
\usepackage[verbose,hypertexnames=false]{hyperref}
\hypersetup{colorlinks=false,allbordercolors=blue,pdfborderstyle={/S/U/W 1}}

\begin{document}

\markboth{Alimov, Saleev}{$J/\psi\gamma$: SPS and DPS}

\catchline{}{}{}{}{}

\title{Associated production of $J/\psi$ mesons and photons in the Parton Reggeization Approach and the double parton scattering model}

\author{Lev Alimov}
\address{Samara National Research University, Moskovskoe Shosse, 34, 443086, Samara, Russia\\
alimov.le@yandex.ru}
\author{Vladimir Saleev}
\address{Samara National Research University, Moskovskoe Shosse, 34, 443086, Samara, Russia}
\address{Joint Institute for Nuclear Research, Dubna, 141980 Russia\\
saleev.vladimir@gmail.com}

\maketitle

\begin{history}
\received{(Day Month Year)}
\revised{(Day Month Year)}
\accepted{(Day Month Year)}
\published{(Day Month Year)}
\end{history}

\begin{abstract}
We study the contribution of double parton scattering (DPS) to the associated production of $J/\psi$ mesons and photons with large transverse momenta in proton–proton collisions. Cross sections are computed within high-energy factorization using the Parton Reggeization Approach (PRA). We used two frameworks for hadronization of the $c\bar c$ pair into charmonium: nonrelativistic QCD (NRQCD) and the improved color evaporation model (ICEM). Hadronization model parameters are fixed using single $J/\psi$ production experimental data from the CMS and ATLAS collaborations at the Large Hadron Collider (LHC). We show that the DPS contribution significantly exceeds the single parton scattering (SPS) contribution and that theoretical predictions are strongly sensitive to the choice of hadronization model. We made predictions for various differential cross sections and correlation spectra for the associated $J/\psi$ and photon production at $\sqrt{s}=13$ TeV.
\end{abstract}


\ccode{PACS Nos.: 14.40.Gx, 13.87.Ce}

\section{Introduction}

The study of the $J/\psi$ mesons and photons associated production in high-energy proton–proton collisions is important both for testing  heavy quarks into quarkonia hadronization models \cite{Mehen:1996vx, Doncheski:1993dj} and for constraining parton distribution functions (PDFs) in the proton\cite{Drees:1991ig}, including transverse-momentum-dependent (TMD) gluon PDFs\cite{ denDunnen:2014kjo}.

Due to the smallness of the QCD coupling constant at the charmonium mass scale, it is able to use perturbative QCD methods to calculate charmonium production cross sections at high energies. In the collinear parton model (CPM), inclusive $J/\psi$ and associated $J/\psi+\gamma$ production cross sections have been calculated up to next-to-leading order\cite{Butenschoen:2012qr, Li:2008ym}. However, those calculations are constrained to large transverse momenta, $p_T\gg m_\psi$, both  for photons and $J/\psi$ mesons. TMD factorization is applicable at small transverse momenta of $J/\psi$, $p_{T\psi}\ll m_\psi$\cite{collins1989factorization}. In this work, we use the high-energy factorization, also referred  as the $k_T$-factorization\cite{collins1991heavy,catani1994high, gribov1983semihard}. As it is was shown in Refs.\cite{Nefedov:2013ywa, Karpishkov:2017kph, Nefedov:2020ugj}, within the gauge-invariant Parton Reggeization Approach (PRA), which is based on high-energy factorization in the multi–Regge kinematics limit, one can describe differential cross sections for heavy-quark, quarkonium, and photon production over the full experimentally probed range of transverse momenta.

The two main frameworks for the nonperturbative transition of a heavy quark–antiquark pair into heavy quarkonium are non-relativistic QCD (NRQCD)\cite{bodwin1995rigorous} and the improved color evaporation model (ICEM)\cite{Ma:2016exq}. In the NRQCD, the $J/\psi$ is formed via intermediate $c\bar c$ states whose contributions scale with different powers of the relative heavy-quark velocity in the bound state. A special case of the NRQCD is the color-singlet model (CSM), which includes only the leading order color-singlet state with the quantum numbers of the final quarkonium, i.e., it neglects relativistic corrections\cite{Baier:1983va, Berger:1980ni}. In the ICEM, charmonium arises from $c\bar c$-pairs with invariant masses in the interval from the charmonium mass up to the threshold for producing a $D$-meson pair, via soft-gluon emission and absorption. In this way, all nonperturbative effects are encoded in a single effective parameter associated with the probability of a  quarkonium production, $\mathcal{F}^C$\cite{Fritzsch:1977ay, Ma:2016exq}.

Despite the existence of extensive experimental data on inclusive production of single $J/\psi$-meson\cite{Brambilla:2010cs,Boer:2024ylx, Zhao:2025sna} and photons\cite{Vogelsang:1997cq, ATLAS:2021mbt, ALICE:2019rtd, CMS:2018qao} from RHIC, Tevatron, and LHC, associated $J/\psi+\gamma$ production remains experimentally unexplored.

In the recent paper\cite{Alimov:2024pqt}, we made predictions for differential cross sections of associated $J/\psi$ and prompt-photon production in the PRA at LHC energies $\sqrt{s}=13–14$ TeV within the single parton scattering (SPS) mechanism, using two hadronization models, CSM and ICEM. It was shown that, at LHC energies, quark–antiquark annihilation and color-octet NRQCD channel contributions are negligible in the low-$p_T$ region for $J/\psi$ mesons, when $p_{T\psi}<20$ GeV. Another important observation was that the ICEM cross section is strongly suppressed compared to the NRQCD prediction for $J/\psi+\gamma$ production, even though NRQCD and ICEM within PRA both well describe single $J/\psi$ and high-$p_T$ photon experimental data.

Given that both experimental results and theory for associated production of heavy quarkonia, $DD$-pairs, and quarkonium and $D-$meson pair indicate a dominant role of double parton scattering (DPS)\cite{Chapon:2020heu}, we have computed differential cross sections for associated $J/\psi$ and photon production within the PRA using NRQCD and ICEM at $\sqrt{s}=13$ TeV in the central $|y_{\gamma, \psi}| < 2$ and the forward $2.0<y_{\gamma, \psi}<4.5$ rapidity regions. Our results confirm the dominant role of the DPS mechnism in the $J/\psi+\gamma$ associated production.

\section{Parton Reggeization approach in SPS model}

Key elements and the current status of the PRA are presented in Refs.\cite{Nefedov:2013ywa, Karpishkov:2017kph, Nefedov:2020ugj}. PRA is a gauge-invariant version of the $k_T$-factorization approach\cite{collins1991heavy,catani1994high,gribov1983semihard}. In general form, the PRA formula for the SPS cross section to produce a particle $A$ reads as a convolution of two unintegrated PDFs (uPDFs) with the hard-scattering cross section for two Reggeized partons:
\begin{equation}
    \sigma^{SPS}(pp\to A X) = \sum\limits_{a,b} \Phi_a\otimes\hat{\sigma}(ab\to AX)\otimes\Phi_b,
    \label{eq:factor}
\end{equation}
where, in our case, $A=J/\psi$ or $\gamma$, $a,b=R, Q$ denote Reggeized gluons and quarks of various flavors in the initial state. The uPDF $\Phi_{a,b}=\Phi_{a,b}(x, t, \mu)$ depends on the longitudinal momentum fraction $x$, the parton transverse-momentum squared $t=\vec q_T^2$, and the hard scale $\mu$. The Reggeized parton scattering cross sections are expressed via gauge-invariant Reggeized amplitudes built using the Feynman rules of the Lipatov’s effective field theory of Reggeized gluons and quarks\cite{lipatov1995gauge,Lipatov:2000se}. In the PRA, it is used the modified Kimber–Martin–Ryskin–Watt (KMRW) model for uPDFs\cite{kimber2001unintegrated, watt2003unintegrated} and all formulae  and derivations are given in Ref.\cite{Nefedov:2020ugj}. Details of calculations for $J/\psi+\gamma$ production in the SPS within the PRA can be found in Ref.\cite{Alimov:2024pqt}. As collinear input to calculate used here uPDFs, we take the MSTW2008lo set~\cite{Martin:2009iq}.

\section{$J/\psi$ production in PRA using NRQCD}

Direct $J/\psi$ production and feeddown production from short-lived excited charmonia ($\psi(2S), \chi_{cJ}$) have been extensively studied in the PRA and the NRQCD\cite{Kniehl:2006sk,nefedov2013charmonium,kniehl2016psi,Saleev:2012hi}, where basic formulae and parameter values are provided.  
Within the NRQCD, the direct charmonium production cross section $\sigma^{SPS}(pp\to {\cal C} X)$ is expanded over Fock states $[n]$ with different orders in the heavy-quark relative velocity:
\begin{equation}
\sigma(pp\to {\cal C} X)=\sum\limits_n \hat{\sigma}(pp\to c
\bar{c}[n] X) \frac{\langle {O}^{\cal C}[n]\rangle}{N_{col}N_{pol}},
\end{equation}
where the sum runs over $n={}^{2S+1}L_J^{(1,8)}$ (spin, color , orbital and total angular momenta) of the $c\bar c$-pair, $N_{col}=2N_c$ for color-singlet states and $N_{col}=N_c^2-1$ for color-octet states, $N_{pol}=2J+1$. The long-distance matrix elements (LDMEs) $\langle {O}^{\cal C}[n]\rangle$ factorize nonperturbative hadronization effects\cite{bodwin1995rigorous}.

Feeddown contribution into $J/\psi$ production involves a kinematic shift in transverse momentum,
$$ {p_{T\psi}}\simeq  \frac{M_{\psi}}{M_{\cal C}} p_{T{\cal C}}  ,$$ 
with ${\cal C}=\psi(2S), \chi_{cJ}$. The contribution of ${\cal C}\to J/\psi X$ to the $J/\psi$ spectrum is then
\begin{align}
    \frac{d\sigma(p_{T\psi})}{dp_{T\psi}}= B({\cal C}\to J/\psi X)\times  \frac{d\sigma(p_{T{\cal C}})}{dp_{T{\cal C}}} \Big|_{p_{T{\cal C}}\to p_{T\psi} \frac{M_{\cal C}}{M_{\psi}}},
\end{align}
where $B({\cal C}\to{J/\psi} X)$ is the relevant branching fraction.

Color-singlet LDMEs $\langle { O}^{\cal C}[n] \rangle$ can be extracted from electromagnetic decay widths or calculated in heavy-quark potential models. Color-octet LDMEs are treated as free parameters. In our work, octet LDMEs are fitted to recent CMS and ATLAS data\cite{CMS:2017dju, ATLAS:2023qnh} on the sum of direct and feeddown $J/\psi$ production. The fit proceeds in two steps. First, using single $\psi(2S)$ data, we fix $\langle { O}^{\psi(2S)}[n] \rangle$, accounting for following suprocesses
\begin{eqnarray}
   && RR\to c\bar{c}[{}^3S_1^{(1)}]g\to \psi(2S)g,\label{eq:decay:psi2s:1}\\
   &&  RR\to c\bar{c}[{}^1S_0^{(8)}, {}^3S_1^{(8)}, {}^3P_J^{(8)}]\to \psi(2S),\label{eq:decay:psi2s:2}
\end{eqnarray}
with $J=0,1,2$. Second, given $\langle { O}^{\psi(2S)}[n] \rangle$, we fix $\langle { O}^{J/\psi}[n] \rangle$ taking into account contributions from next suprocesses
\begin{eqnarray*}
  &&  RR\to c\bar{c}[{}^3S_1^{(1)}]g\to J/\psi g,\\
   && RR\to c\bar{c}[{}^1S_0^{(8)}, {}^3S_1^{(8)}, {}^3P_J^{(8)}]\to \psi(2S)\to J/\psi, \\
   && RR\to c\bar{c}[{}^1S_0^{(8)},{}^3S_1^{(8)},{}^3P_J^{(1,8)}]\to\chi_{cJ}\to J/\psi.
\end{eqnarray*}
The color-singlet and fitted color-octet LDMEs used for calculation of the $J/\psi+\gamma$ cross sections are summarized in Table~\ref{tab:NME}. Masses employed are ${m_{J/\psi}=3.096}$ GeV, ${m_{\chi_{c0}}=3.415}$ GeV, ${m_{\chi_{c1}}=3.511}$ GeV, ${m_{\chi_{c2}}=3.556}$ GeV, ${m_{\psi(2S)}=3.686}$ GeV. The quark–antiquark annihilation contribution is negligible for $J/\psi$ production at LHC energies within NRQCD\cite{Kniehl:2006sk, Nefedov:2013hya}. Used squared amplitudes for single-charmonium production via Reggeized gluon-gluon fusion are taken from Ref.\cite{Kniehl:2006sk}. 

\section{$J/\psi$ production in PRA using ICEM}

In ICEM, the prompt $J/\psi$ production cross section $\sigma^{SPS}(pp\to J/\psi X)$ is obtained by integrating the $c\bar c$-pair production cross section over the pair invariant mass $M$ from $m_{J/\psi}$ up to the $D\bar D$ threshold:
\begin{align}
\sigma^{SPS}(pp\to J/\psi X)\simeq \mathcal{F}^{J/\psi} \int\limits_{m_{J/\psi}}^{2m_D} dM \frac{d\hat\sigma(pp\to c\bar{c})}{dM}{\Big |}_{p = (M/m_{\psi}) p_{\psi}}.
\end{align}
Following Ref.\cite{Chernyshev:2022gek}, we include only the gluon-gluon fusion partonic subprocess in calculations, since quark–antiquark annihilation contribution is small at high energies. The ICEM parameter is set to be equal $\mathcal{F}^{J/\psi}=0.02$\cite{Chernyshev:2022gek}, and we use ${m_c=1.3}$~GeV, ${m_D=1.87}$~GeV.

To calculate differential cross sections of $J/\psi$ and  $J/\psi+\gamma$ production in the PRA using the ICEM, we use the parton-level Monte Carlo generator KaTie\cite{van2018katie}. It has been verified\cite{Nefedov:2013ywa} that amplitudes for processes with Reggeized partons numerically computed using the AvhLib library~\cite{vanHameren:2012if} via spinor-amplitude techniques and BCFW recursion\cite{Britto:2005fq} agree with tree-level amplitudes obtained from Lipatov’s effective theory\cite{lipatov1995gauge}.

\section{Prompt photon production in PRA}

Prompt photon production in the PRA was first studied in Refs.\cite{Saleev:2008rn,Saleev:2011zz}. As it was shown recently\cite{Chernyshev:2024qvq}, in the calculation at leading order (LO) in the strong coupling constant, with next-to-leading order (NLO) corrections from additional parton emission and double-counting subtractions between LO subprocess $Q\bar Q\to \gamma$ and NLO subprocesses $Q\bar Q\to \gamma g$ and $Q R\to \gamma q$, a good phenomenological approximation is to take only the Compton-like scattering of a gluon on a quark (or an antiquark):
\begin{align}
    RQ(\bar Q)\to \gamma q (\bar q).
    \label{eq:gamma_subproc}
\end{align}
Both theory and experiment includes photon isolation with a cone parameter $R = \sqrt{\Delta\phi^2+\Delta\eta^2} > 0.4$, which is important to suppress double counting and uncertainties from fragmentation photons.
For the subprocess in Eq.~\eqref{eq:gamma_subproc} we use the analytic Reggeized amplitude from Ref.\cite{Saleev:2011zz}.

\section{Double parton scattering}

Assuming that two hard subprocesses are independent, the DPS contribution to the associated $J/\psi+\gamma$ cross section is given by the standard "pocket formula":
\begin{align}
    \sigma^{DPS}(pp\to J/\psi\gamma)=\frac{\sigma^{SPS}(pp\to J/\psi X)\times\sigma^{SPS}(pp\to \gamma X)}{\sigma^{eff}},
\end{align}
where $\sigma^{eff}$ is an effective cross section that controles the DPS contribution value. The SPS components $\sigma^{SPS}(pp\to J/\psi X),~\sigma^{SPS}(pp\to \gamma X)$ are computed within the PRA using  Eq.~\eqref{eq:factor}.

We take $\sigma_{eff}=11.0\pm 0.2$ mb, as it was obtained by describing data on associated production of $J/\psi J/\psi$, $J/\psi \psi(2S)$, $J/\psi \Upsilon$, $\Upsilon\Upsilon$, $J/\psi Z$, and $J/\psi W$\cite{Chernyshev:2024nei, Chernyshev:2023qea, Chernyshev:2023kzk, Chernyshev:2022gek} within the PRA using the DPS model.

\section{Results}

\begin{figure}[ht]
\centerline{\includegraphics[width=\textwidth]{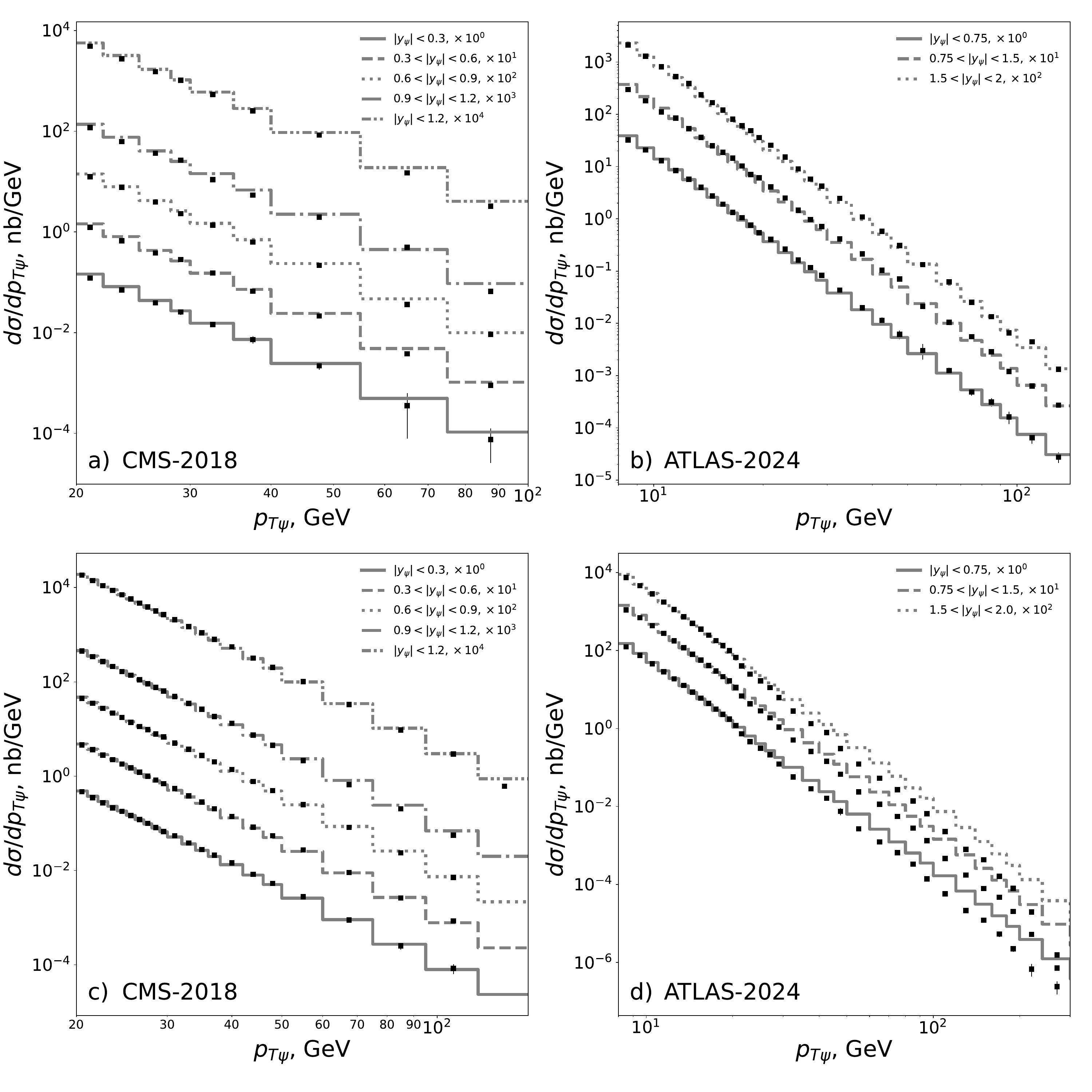}}
\vspace*{8pt}
\caption{Differential cross sections as functions of $J/\psi$ or $\psi(2S)$ transverse momenta: panels a) and b) -- $\psi(2S)$ production , panels c) and d) -- $J/\psi$ production. The data are from CMS\cite{CMS:2017dju} and  ATLAS\cite{ATLAS:2023qnh} collaborations.}
\label{fig1}
\alttext{LDME fit results for $p_T$ spectra of $J/\psi$ and $\psi(2S)$, compared to CMS and ATLAS data.}
\end{figure}

First, within the PRA using the SPS we fit CMS\cite{CMS:2017dju} and ATLAS\cite{ATLAS:2023qnh} $J/\psi$ production data to obtain charmonium octet LDMEs, see Table~\ref{tab:NME}. Some LDMEs are set to zero because, in the kinematic regions considered, the $RR\to c\bar{c}[{}^1S_0^{(8)}]$, $RR\to c\bar{c}[{}^3S_1^{(8)}]$, and $RR\to c\bar{c}[{}^3P_J^{(8)}]$ subprocess contributions are linearly dependent. In such situations, dimensionality-reduction techniques are often used, e.g. introducing combinations like ${M_r=\langle O^\mathcal{C} [{}^1S_0^{(8)}]\rangle + \frac{r}{m_c^2}\langle O^\mathcal{C} [{}^3P_J^{(8)}]\rangle}$. In our data set, with $d.o.f.=328$, the $\chi^2$ does not increase significantly when the parameter space dimension grows, and to regularize it suffices to set linearly dependent terms to zero. Table~\ref{tab:NME} also lists LDME fit uncertainties from $\chi^2$ variation on $\pm 1$. The description of $J/\psi$ and $\psi(2S)$ transverse momentum spectra~\cite{CMS:2017dju, ATLAS:2023qnh} with the fitted LDMEs is shown in Fig.~\ref{fig1}. For the $\psi(2S)$ production, the good agreement with the experimental data is observed, see panels a) and b) in  Fig.~\ref{fig1}. For single $J/\psi$ production cross sections, agreement is observed only at small $p_{T\psi}$. Therefore our predictions for $J/\psi+\gamma$ production are restricted to $p_{T\psi}<20$ GeV to exclude a kinematic region where predictions would be unreliable.

\begin{table}[ht]
\tbl{Color-singlet and color-octet LDMEs\label{tab:NME}}
{\tabcolsep13pt\begin{tabular}{@{}llll@{}}
\toprule

$\langle { O}^{\psi(2S)}[{}^{3}S_1^{(1)}] \rangle/\text{GeV}^3$ &
$6.5 \cdot 10^{-1}$ &
$\langle { O}^{\psi(2S)}[{}^{3}S_1^{(8)}] \rangle/\text{GeV}^3$ &
$9.4^{+0.2}_{-0.2} \cdot 10^{-4}$ \\
$\langle { O}^{\psi(2S)}[{}^{1}S_0^{(8)}] \rangle/\text{GeV}^3$ &
$0.0^{+5.7}_{-5.8} \cdot 10^{-4}$ &
$\langle { O}^{\psi(2S)}[{}^{3}P_0^{(8)}] \rangle/\text{GeV}^5$ &
$1.01^{+0.03}_{-0.03} \cdot 10^{-2}$ \\
\colrule
$\langle { O}^{J/\psi}[{}^{3}S_1^{(1)}] \rangle/\text{GeV}^3$ &
$1.3$ & 
$\langle { O}^{\chi_{cJ}}[{}^{3}P_0^{(1)}] \rangle/\text{GeV}^5$ &
$8.9 \cdot 10^{-2}$ \\
$\langle { O}^{J/\psi}[{}^{1}S_0^{(8)}] \rangle/\text{GeV}^3$ &
$4.54^{+0.07}_{-0.07} \cdot 10^{-2}$ &
$\langle { O}^{\chi_{cJ}}[{}^{1}S_0^{(8)}] \rangle/\text{GeV}^3$ &
$0.0^{+0.7}_{-1.2} \cdot 10^{-3}$ \\
$\langle { O}^{J/\psi}[{}^{3}S_1^{(8)}] \rangle/\text{GeV}^3$ &
$0.0^{+0.9}_{-2.2} \cdot 10^{-5}$&
$\langle { O}^{\chi_{cJ}}[{}^{3}S_1^{(8)}] \rangle/\text{GeV}^3$ &
$1.59^{+0.02}_{-0.02} \cdot 10^{-3}$\\
$\langle { O}^{J/\psi}[{}^{3}P_0^{(8)}] \rangle/\text{GeV}^5$ &
$0.0^{+2.7}_{-3.2} \cdot 10^{-4}$&
$\langle { O}^{\chi_{cJ}}[{}^{3}P_0^{(8)}] \rangle/\text{GeV}^5$ &
$0.0^{+4.5}_{-5.8} \cdot 10^{-3}$\\
\botrule
\end{tabular}}
\end{table}

\begin{figure}[ht]
\centerline{\includegraphics[trim={0 41.5cm 0 0},clip,width=\textwidth]{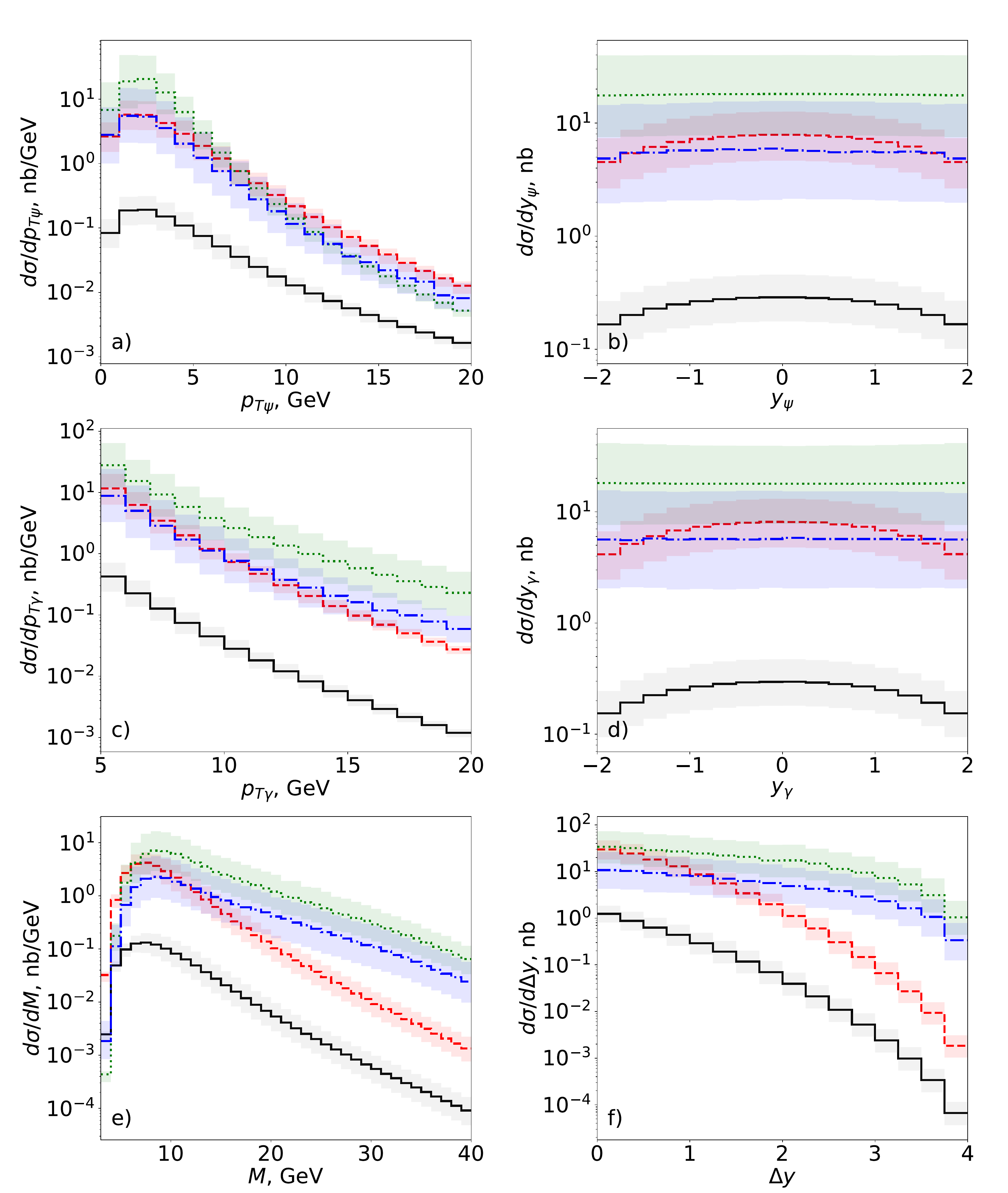}}
\vspace*{8pt}
\caption{Differential cross sections for associated $J/\psi+\gamma$ production in the PRA at $\sqrt{s}=13$ TeV in the central rapidity region $|y_\psi|, |y_{\gamma}| < 2$: a) as a function of $J/\psi$ transverse momentum, b)  $J/\psi$ rapidity, c) photon transverse momentum, d) photon rapidity, e) ${J/\psi+\gamma}$ pair invariant mass, f)  rapidity difference. Contribution of the DPS using the NRQCD -- dotted histogram, the DPS using the ICEM -- dot–dashed histogram, the SPS using the  NRQCD -- dashed histogram, the SPS using the ICEM -- solid histogram. Blue and gray bands indicate the effect of hard scale $\mu$ variation by a factor of 2.}
\label{fig2}
\alttext{SPS and DPS comparisons for $J/\psi\gamma$ at $\sqrt{s}=13$ TeV across $p_T$, rapidity, and angular observables.}
\end{figure}

Using the obtained LDMEs in the PRA using the NRQCD and the ICEM hadronization models at energy $\sqrt{s}=13$ TeV in the region $|y_\psi|, |y_\gamma|<3$, $p_{T\psi}<20$ GeV, $p_{T\gamma}>5$ GeV, we calculate the DPS contribution to the different differential cross sections. In the Fig.~\ref{fig2}, they are shown as functions of transverse momenta and rapidities of $J/\psi$-meson and photon, the rapidity difference $\Delta y=|y_\psi-y_\gamma|$, and the pair invariant mass $M$.
In the Fig.~\ref{fig3}, they are shown as functions of the azimuthal angle difference $\Delta\phi = |\phi_{\psi}-\phi_{\gamma}|$, the pair rapidity $Y=Y_{\psi\gamma}$, the pair transverse momentum $p_T=|\mathbf{p}_{T\psi}+\mathbf{p}_{T\gamma}|$, and the transverse asymmetry $\mathcal{A}_T=(|\mathbf{p}_{T\psi}|-|\mathbf{p}_{T\gamma}|)/(|\mathbf{p}_{T\psi}|+|\mathbf{p}_{T\gamma}|)$.

By the same way, we calculate SPS and DPS contributions for the $J/\psi+\gamma$ production differential cross sections in forward rapidity region of $J/\psi$ and photon, $2.0<y_\psi,y_\gamma<4.5$,  at $p_{T\psi}<20$ GeV, $p_{T\gamma}>5$ GeV. In the Fig.~\ref{fig4}, they are shown as functions of transverse momenta and rapidities of $J/\psi$-meson and photon, the rapidity difference $\Delta y$, and the pair invariant mass $M$.
In the Fig.~\ref{fig5}, they are shown as functions of the azimuthal angle difference $\Delta\phi$, the pair rapidity $Y$, the pair transverse momentum $p_T$, and the transverse asymmetry $\mathcal{A}_T$.
The shaded bands in Figs.~\ref{fig2}--\ref{fig5} indicate the theoretical uncertainty due to the choice of hard scale, obtained variation by factor of $2$.

The results of calculations of the differential cross sections in the PRA using the ICEM and the NRQCD
 in the SPS model , as it was calculated in Ref.~\cite{Alimov:2024pqt},  are also shown for comparison in Figs.~\ref{fig2}--\ref{fig5} .

Thus, we have obtained that the DPS contribution significantly exceeds the SPS contribution in the associated production of $J/\psi+\gamma$. The same as in the SPS model, the DPS contribution computed in the PRA using the NRQCD is larger than the DPS contribution computed in the PRA using the  ICEM.

\section{Conclusions}

Within PRA, we have shown that the contribution of the DPS production mechanism dominates over the SPS contribution in associated $J/\psi+\gamma$ production, independently of the hadronization model used. At the same time, both SPS\cite{Alimov:2024pqt} and presented here DPS calculations shown that the NRQCD model of hadronization yields substantially larger $J/\psi+\gamma$ production cross sections than the ICEM.

\section{Acknowledgments}

The work is supported by the Foundation for the Advancement of Theoretical Physics and Mathematics BASIS, grant No. 24–1–1–16–5 and by the grant of the Ministry of Science and Higher Education of the Russian Federation, No. FSSS 2025–0003.

\bibliographystyle{ws-mpla}
\bibliography{biblio}
\newpage

\begin{figure}[ht]
\centerline{\includegraphics[trim={0 0 0 60cm},clip,width=\textwidth]{LHC13_cntr.pdf}}
\vspace*{8pt}
\caption{
Differential cross sections for associated $J/\psi+\gamma$ production in the PRA at $\sqrt{s}=13$ TeV in the central rapidity region $|y_\psi|, |y_{\gamma}| < 2$: a) as a function of azimuthal angle difference, b) pair rapidity, c) pair transverse momentum, d) transverse-momentum asymmetry. Contribution of the DPS using the NRQCD -- dotted histogram, the DPS using the ICEM -- dot–dashed histogram, the SPS using the NRQCD -- dashed histogram, the SPS using the ICEM -- solid histogram. Blue and gray bands indicate the effect of hard scale $\mu$ variation   by a factor of 2.}
\label{fig3}
\alttext{SPS and DPS comparisons for $J/\psi+\gamma$ at $\sqrt{s}=13$ TeV across pair mass, $p_T$ asymmetry, pair $p_T$, and pair rapidity.}
\end{figure}

\begin{figure}[ht]
\centerline{\includegraphics[trim={0 41.5cm 0 0},clip,width=\textwidth]{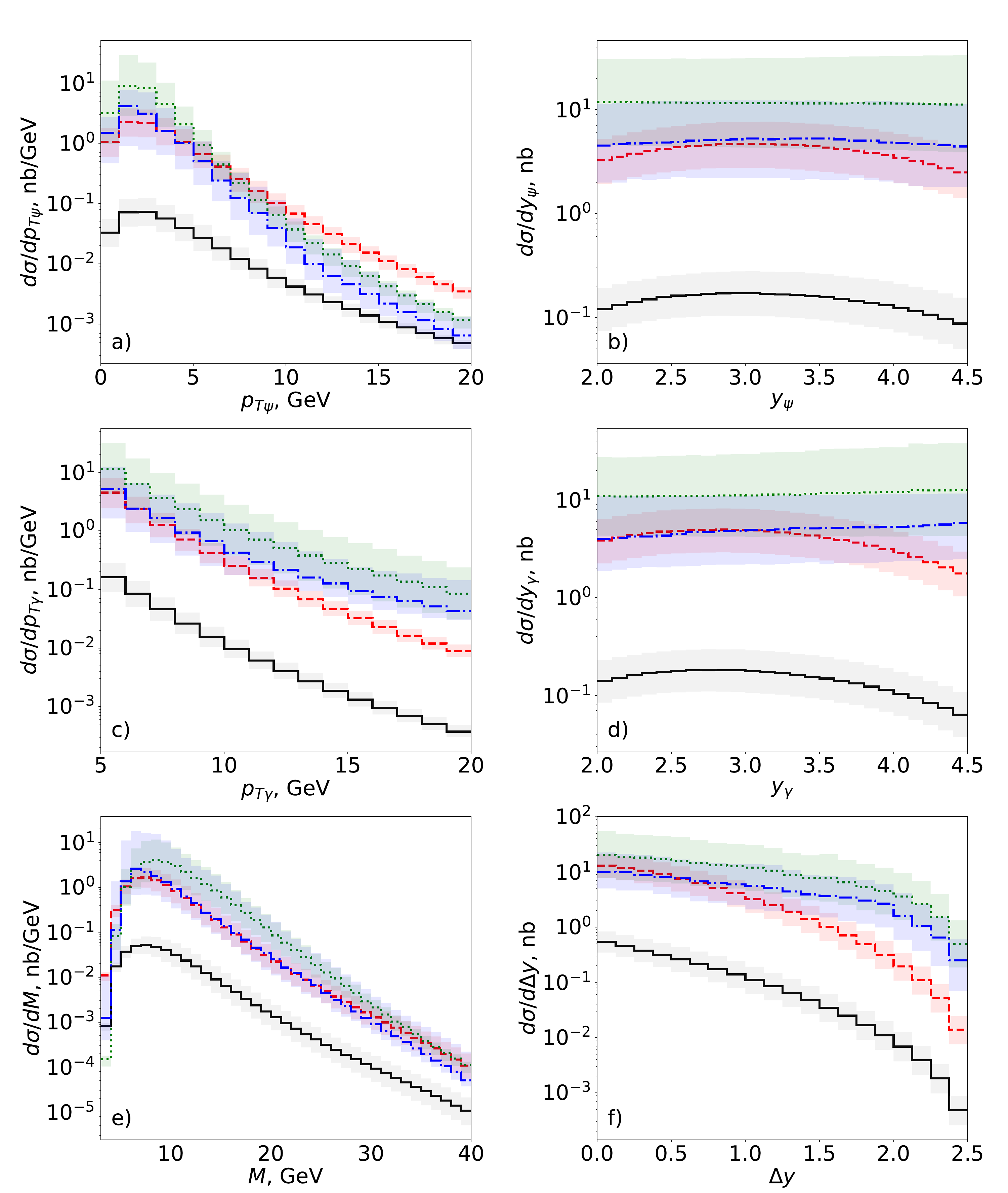}}
\vspace*{8pt}
\caption{Differential cross sections for associated $J/\psi+\gamma$ production in the PRA at $\sqrt{s}=13$ TeV in the forward rapidity region $2.0<y_\psi, y_{\gamma} < 4.5$:  a) as a function of $J/\psi$ transverse momentum, b)  $J/\psi$ rapidity, c) photon transverse momentum, d) photon rapidity, e) ${J/\psi+\gamma}$ pair invariant mass, f)  rapidity difference. Contribution of the DPS using the NRQCD -- dotted histogram, the DPS using the ICEM -- dot–dashed histogram, the SPS using the  NRQCD -- dashed histogram, the SPS using the ICEM -- solid histogram. Blue and gray bands indicate the effect of hard scale $\mu$ variation by a factor of 2.}
\label{fig4}
\alttext{SPS and DPS comparisons for $J/\psi\gamma$ at $\sqrt{s}=13$ TeV across $p_T$, rapidity, and angular observables.}
\end{figure}

\begin{figure}[ht]
\centerline{\includegraphics[trim={0 0 0 60cm},clip,width=\textwidth]{LHC13_frwd.pdf}}
\vspace*{8pt}
\caption{
Differential cross sections for associated $J/\psi+\gamma$ production in the PRA at $\sqrt{s}=13$ TeV in the forward rapidity region $2.0<y_\psi, y_{\gamma} < 4.5$:  a) as a function of azimuthal angle difference,  b) pair rapidity, c) pair transverse momentum, d) transverse-momentum asymmetry. Contribution of the DPS using the NRQCD -- dotted histogram, the DPS using the ICEM -- dot–dashed histogram, the SPS using the NRQCD -- dashed histogram, the SPS using the ICEM -- solid histogram. Blue and gray bands indicate the effect of hard scale $\mu$ variation   by a factor of 2.}
\label{fig5}
\alttext{SPS and DPS comparisons for $J/\psi+\gamma$ at $\sqrt{s}=13$ TeV across pair mass, $p_T$ asymmetry, pair $p_T$, and pair rapidity.}
\end{figure}

\end{document}